\documentclass[english,aps,superscriptaddress,preprintnumbers]{revtex4}
\usepackage[T1]{fontenc}
\usepackage[latin9]{inputenc}
\usepackage{mathrsfs}
\usepackage{amsmath}
\usepackage{amssymb}
\usepackage{graphicx}
\usepackage{esint}

\makeatletter
\@ifundefined{textcolor}{}
{%
 \definecolor{BLACK}{gray}{0}
 \definecolor{WHITE}{gray}{1}
 \definecolor{RED}{rgb}{1,0,0}
 \definecolor{GREEN}{rgb}{0,1,0}
 \definecolor{BLUE}{rgb}{0,0,1}
 \definecolor{CYAN}{cmyk}{1,0,0,0}
 \definecolor{MAGENTA}{cmyk}{0,1,0,0}
 \definecolor{YELLOW}{cmyk}{0,0,1,0}
}

\makeatother

\usepackage{babel}

\begin{document}

\title{Polarization of massive fermions in a vortical fluid}

\author{Ren-hong Fang}
\affiliation{Interdisciplinary Center for Theoretical Study and Department of
Modern Physics, University of Science and Technology of China, Hefei,
Anhui 230026, China}

\author{Long-gang Pang}
\affiliation{Frankfurt Institute for Advanced Studies, Ruth-Moufang-Strasse 1,
60438 Frankfurt am Main, Germany }

\author{Qun Wang}
\affiliation{Interdisciplinary Center for Theoretical Study and Department of
Modern Physics, University of Science and Technology of China, Hefei,
Anhui 230026, China}

\author{Xin-nian Wang}
\affiliation{Key Laboratory of Quark and Lepton Physics (MOE) and Institute of
Particle Physics, Central China Normal University, Wuhan, 430079,
China}
\affiliation{Nuclear Science Division, MS 70R0319, Lawrence Berkeley National
Laboratory, Berkeley, California 94720}

\begin{abstract}
Fermions become polarized in a vortical fluid due to spin-vorticity coupling. 
Such a polarization can be calculated from the Wigner function in a quantum kinetic approach. 
Extending previous results for chiral fermions, we derive the Wigner function for massive 
fermions up to the next-to-leading order in spatial gradient expansion. The polarization density of fermions can be calculated from the axial vector component of the Wigner function and is found to be proportional 
to the local vorticity $\boldsymbol\omega$. 
The polarizations per particle for fermions and anti-fermions decrease with the chemical potential and 
increase with energy (mass). Both quantities approach the asymptotic value $\hbar\boldsymbol\omega/4$ 
in the large energy (mass) limit. The polarization per particle for fermions 
is always smaller than that for anti-fermions, whose ratio of fermions 
to anti-fermions also decreases with the chemical potential. 
The polarization per particle on the Cooper-Frye freeze-out hyper-surface 
can also be formulated and is consistent with the previous result of Becattini et al..
\end{abstract}

\preprint{\hfill {\small {ICTS-USTC-16-05}}}

\maketitle

\section{Introduction}

In non-central high-energy heavy-ion collisions, the large orbital angular momentum present 
in the colliding system can lead to non-vanishing local vorticity in the hot and 
dense fluid \cite{Liang:2004ph,Liang:2004xn,Becattini:2007sr,Betz:2007kg,Gao:2007bc,Huang:2011ru}.
The vorticity induced by global orbital angular momentum in the fluid can be considered 
as local rotational motion of particles \cite{Becattini:2007sr,Betz:2007kg,Jiang:2016woz,Deng:2016gyh}.
It is closely related to the rapidity dependence of the $v_{1}$ flow and shear of the longitudinal flow velocity inside the reaction plane \cite{Gao:2007bc,Csernai:2011gg,Wang:2013xtp}.

As a result of spin-orbital coupling, quarks and anti-quarks can become polarized 
along the normal direction of the reaction plane \cite{Liang:2004ph,Liang:2004xn,Gao:2007bc}.
Through hadronization of polarized quarks and anti-quarks, hyperons can also be polarized 
in the same direction in the final state \cite{Liang:2004ph,Liang:2004xn,Becattini:2013vja}. 
Measurements of such global hyperon polarization is feasible through the parity-violating 
decay of hyperons \cite{Abelev:2007zk,Deng:2014eha}. Such measurements will shed light on properties of the vortical structures of the strongly coupled quark-gluon plasma (sQGP) in high-energy heavy-ion collisions. 

Quark and anti-quark polarization in a vortical fluid is also closely 
related to the Chiral Magnetic and Vortical Effects 
\cite{Kharzeev:2007jp,Fukushima:2008xe,Son:2009tf,Kharzeev:2010gr,Pu:2010as,Gao:2012ix}. 
From the solutions of Wigner functions for chiral or massless fermions in a quantum kinetic approach
one can derive the axial current $j_{5}^{\mu}=\rho_{5}u^{\mu}+\xi_{5}\omega^{\mu}+\xi_{5}^{B}B^{\mu}$,
where $\rho_{5}$ is the axial charge density, $u^{\mu}$ is the
fluid velocity, 
$\omega^{\mu}\equiv\frac{1}{2}\epsilon^{\mu\sigma\alpha\beta}u_{\sigma}\partial_{\alpha}u_{\beta}$
is the vorticity 4-vector, and 
$B^{\mu}=\frac{1}{2}\epsilon^{\mu\nu\lambda\sigma}u_{\nu}F_{\lambda\sigma}$
is the 4-vector of the magnetic field with $F_{\lambda\rho}$ being 
the strength tensor of the electromagnetic field. The coefficients
$\xi_{5}$ and $\xi_{5}^{B}$ are all functions of temperatures and chemical potentials
$\mu$ and $\mu_{5}$ \cite{Gao:2012ix}. In a three-flavor
quark matter with u, d and s quarks and their anti-quarks,
$\xi_{5}^{B}=0$. In other words, the axial current in a three-flavor quark matter is blind to the
magnetic field and solely induced by the vorticity. Such an axial
current leads to the Local Polarization Effect \cite{Gao:2012ix}
which is also connected to the spin-vorticity coupling for chiral
or massless fermions \cite{Gao:2015zka}. 

In this paper, we will extend our Wigner function method for massless
fermions to massive ones and formulate the polarization of massive
fermions induced by vorticity. In Section \ref{sec:wigner-mass},
we will give a brief introduction to the Wigner function method and derive
the equations for the Wigner function components for massive fermions
based on Ref. \cite{Elze:1986qd,Vasak:1987um}. The Wigner function
components can be determined perturbatively by gradient expansion.
In Section \ref{sec:lead-order}, we will derive the Wigner function
at the leading order by definition. Using the projection method we
can extract each component of the Wigner function at the leading order.
We will propose the first order solution for the axial vector component
in Section \ref{sec:first-order} by extending the solution for massless
fermions. In Section \ref{sec:spin-tensor}, we will show that the
axial vector component can be regarded as the spin density in phase
space. We can obtain the polarization density after completion of
momentum integration of the axial vector component in Section \ref{sec:polar}.
We will also formulate the fermion polarization on the freezeout hypersurface
by extending the Cooper-Frye formula. We will give a summary of the
results in the final section. 

We adopt the same sign conventions for fermion charge $Q$ as in Refs.
\cite{Vasak:1987um,Gao:2012ix,Chen:2012ca,Gao:2015zka}, and the same
sign convention for the axial vector $A^{\mu}\sim\left\langle \bar{\psi}\gamma^{\mu}\gamma^{5}\psi\right\rangle $
as in Resf. \cite{Gao:2012ix,Chen:2012ca,Gao:2015zka} but different
sign convention from Ref. \cite{Vasak:1987um}.

\section{Wigner function for massive fermions }

\label{sec:wigner-mass} In this section we will give a brief introduction
to the Wigner function and its kinetic equation for massive fermions based on Refs. \cite{Elze:1986qd,Vasak:1987um}.
There are also other earlier works in the literature along this line
\cite{Zhuang:1995pd,Ochs:1998qj}. In a background electromagnetic
field, the quantum mechanical analogue of a classical phase-space
distribution for fermions is the gauge invariant Wigner function $W_{\alpha\beta}(x,p)$
defined by 
\begin{equation}
W_{\alpha\beta}(x,p)=\int\frac{d^{4}y}{(2\pi)^{4}}e^{-ip\cdot y}\left\langle \bar{\psi}_{\beta}(x+\frac{1}{2}y)PU(G,x+\frac{1}{2}y,x-\frac{1}{2}y)\psi_{\alpha}(x-\frac{1}{2}y)\right\rangle ,\label{eq:wigner-def-1}
\end{equation}
where $\psi_{\alpha}$ and $\bar{\psi}_{\beta}$ are fermionic quantum
fields, $\langle \hat{O}\rangle $ denotes the grand canonical
ensemble averaging and normal ordering, $x=(x_{0},\mathbf{x})$ and
$p=(p_{0},\mathbf{p})$ are time-space and energy-momentum 4-vectors
respectively, and the gauge link $PU(G,x_{1},x_{2})$ is to ensure
the gauge invariance of the Wigner function and given by 
\begin{equation}
PU(G,x+\frac{1}{2}y,x-\frac{1}{2}y)=P\exp\left[-iQy^{\mu}\int_{0}^{1}dsG_{\mu}(x-\frac{1}{2}y+sy)\right],\label{eq:phase}
\end{equation}
where $G^{\mu}$ is the gauge potential of the classical electromagnetic field. 

The Wigner function in (\ref{eq:wigner-def-1}) satisfies the following
equation of motion, 
\begin{equation}
(\gamma_{\mu}K^{\mu}-m)W(x,p)=0,\label{eq:wigner-dirac}
\end{equation}
where the operator $K^{\mu}$ is given by 
\begin{equation}
K^{\mu}=p_\mathrm{W}^{\mu}+i\hbar\frac{1}{2}\nabla^{\mu},
\end{equation}
with 
\begin{eqnarray}
p_\mathrm{W}^{\mu} & = & p^{\mu}-\hbar\frac{1}{2} Q j_{1}(\Delta )F^{\mu\nu}\partial_{p\,\nu},\nonumber \\
\nabla^{\mu} & = & \partial_{x}^{\mu}-Qj_{0}(\Delta)F^{\mu\nu}\partial_{p\,\nu},
\end{eqnarray}
where we have used $\Delta \equiv \frac{1}{2}\hbar\partial_{p}\cdot\partial_{x}$ 
with the operator $\partial_{x}$ in $\Delta$ acting only on the strength tensor $F^{\mu\nu}$, and 
$j_0(x)=\mathrm{sin}(x)/x$ and $j_1(x)=(\mathrm{sin}(x)-x\mathrm{cos}(x))/x^2$ are spherical Bessel functions. If $F^{\mu\nu}$ is
a constant we have simpler forms of these operators 
\begin{eqnarray}
p_\mathrm{W}^{\mu} & = & p^{\mu},\nonumber \\
\nabla^{\mu} & = & \partial_{x}^{\mu}-QF^{\mu\nu}\partial_{p\,\nu}.\label{eq:constant-field}
\end{eqnarray}
The Wigner function is a $4\times4$ matrix in Dirac indices and can
be decomposed into 16 independent generators of Clifford algebra,
\begin{equation}
W=\frac{1}{4}\left[F+i\gamma^{5}P+\gamma^{\mu}V_{\mu}+\gamma^{5}\gamma^{\mu}A_{\mu}+\frac{1}{2}\sigma^{\mu\nu}S_{\mu\nu}\right],\label{eq:wigner-decomp}
\end{equation}
where the generators of Clifford algebra are 
\begin{equation}
\Gamma_{i}=1,\gamma^{5}=i\gamma^{0}\gamma^{1}\gamma^{2}\gamma^{3},\gamma^{\mu},\gamma^{5}\gamma^{\mu},\sigma^{\mu\nu}=\frac{i}{2}[\gamma^{\mu},\gamma^{\nu}],
\end{equation}
corresponding to the scalar, pseudoscalar, vector, axial vector and
tensor components respectively. The coefficients in the decomposition
(\ref{eq:wigner-decomp}) can be obtained by projection of corresponding
Dirac matrices on the Wigner function and taking traces, 
\begin{eqnarray}
F & = & \mathrm{Tr}[W],\nonumber \\
P & = & -i\mathrm{Tr}[\gamma^{5}W],\nonumber \\
V^{\mu} & = & \mathrm{Tr}[\gamma^{\mu}W],\nonumber \\
A^{\mu} & = & \mathrm{Tr}[\gamma^{\mu}\gamma^{5}W],\nonumber \\
S^{\mu\nu} & = & \mathrm{Tr}[\sigma^{\mu\nu}W].\label{eq:fpvas}
\end{eqnarray}
Substituting Eq. (\ref{eq:wigner-decomp}) into Eq. (\ref{eq:wigner-dirac})
with (\ref{eq:constant-field}) and comparing common terms in the
basis of Clifford algebra, we obtain the following system of equations,
\begin{eqnarray}
K^{\mu}V_{\mu}-mF & = & 0,\nonumber \\
K^{\mu}A_{\mu}+imP & = & 0,\nonumber \\
K_{\mu}F+iK^{\nu}S_{\nu\mu}-mV_{\mu} & = & 0,\nonumber \\
iK_{\mu}P+\frac{1}{2}\epsilon_{\mu\beta\nu\sigma}K^{\beta}S^{\nu\sigma}+mA_{\mu} & = & 0,\nonumber \\
-i(K_{\mu}V_{\nu}-K_{\nu}V_{\mu})-\epsilon_{\mu\nu\alpha\beta}K^{\alpha}A^{\beta}-mS_{\mu\nu} & = & 0.
\end{eqnarray}
The real parts of the above equations are 
\begin{eqnarray}
p^{\mu}V_{\mu}-mF & = & 0,\nonumber \\
\frac{1}{2}\hbar\nabla^{\mu}A_{\mu}+mP & = & 0,\nonumber \\
p_{\mu}F-\frac{1}{2}\hbar\nabla^{\nu}S_{\nu\mu}-mV_{\mu} & = & 0,\nonumber \\
-\frac{1}{2}\hbar\nabla_{\mu}P+\frac{1}{2}\epsilon_{\mu\beta\nu\sigma}p^{\beta}S^{\nu\sigma}+mA_{\mu} & = & 0,\nonumber \\
\frac{1}{2}\hbar(\nabla_{\mu}V_{\nu}-\nabla_{\nu}V_{\mu})-\epsilon_{\mu\nu\alpha\beta}p^{\alpha}A^{\beta}-mS_{\mu\nu} & = & 0.\label{eq:real}
\end{eqnarray}
The imaginary parts are 
\begin{eqnarray}
\hbar\nabla^{\mu}V_{\mu} & = & 0,\nonumber \\
p^{\mu}A_{\mu} & = & 0,\nonumber \\
\frac{1}{2}\hbar\nabla_{\mu}F+p^{\nu}S_{\nu\mu} & = & 0,\nonumber \\
p_{\mu}P+\frac{1}{4}\hbar\epsilon_{\mu\beta\nu\sigma}\nabla^{\beta}S^{\nu\sigma} & = & 0,\nonumber \\
(p_{\mu}V_{\nu}-p_{\nu}V_{\mu})+\frac{1}{2}\hbar\epsilon_{\mu\nu\alpha\beta}\nabla^{\alpha}A^{\beta} & = & 0.\label{eq:im}
\end{eqnarray}

From the 3rd and the 5th line of the imaginary part equations (\ref{eq:im}) we obtain,
\begin{eqnarray}
p\cdot\nabla F & = & 0,
\end{eqnarray}
 and
\begin{eqnarray}
\hbar(\nabla^{\lambda}A^{\rho}-\nabla^{\rho}A^{\lambda})-2\epsilon^{\mu\nu\lambda\rho}p_{\mu}V_{\nu} & = & 0,
\end{eqnarray}
respectively, where we have multiplied $\epsilon^{\mu\nu\lambda\rho}$ to the equation
and used $\epsilon^{\mu\nu\lambda\rho}\epsilon_{\mu\nu\alpha\beta}=-2(\delta_{\alpha}^{\lambda}\delta_{\beta}^{\rho}-\delta_{\beta}^{\lambda}\delta_{\alpha}^{\rho})$.
Taking contraction of the above equation with $p^{\lambda}$, we obtain
\begin{eqnarray}
p\cdot\nabla A^{\rho} & = & p_{\lambda}\nabla^{\rho}A^{\lambda}=QF^{\rho\xi}A_{\xi},
\end{eqnarray}
where we have used $p^{\mu}A_{\mu}=0$ from the 2nd line of Eqs. (\ref{eq:im}). 

From the 1st and 3rd lines of real part equations (\ref{eq:real}),
we obtain 
\begin{eqnarray}
(p^{2}-m^{2})F & = & \frac{1}{2}\hbar p^{\mu}\nabla^{\nu}S_{\nu\mu}\approx\frac{1}{2}\hbar QF^{\mu\nu}S_{\mu\nu},\label{eq:onshell-f}
\end{eqnarray}
where we have neglected the second order term $\hbar\nabla^{\nu}(p^{\mu}S_{\nu\mu})\sim\hbar^{2}$.
Inserting the 5th line into the 4th line in Eqs. (\ref{eq:real})
and neglecting the second order term $\hbar\nabla_{\mu}P\sim\hbar^{2}$,
we obtain 
\begin{eqnarray}
(p^{2}-m^{2})A_{\mu} & = & \frac{1}{2}\hbar\epsilon_{\mu\beta\nu\sigma}p^{\beta}\nabla^{\nu}V^{\sigma}\nonumber \\
 & = & -\frac{1}{2}\hbar Q\epsilon_{\mu\beta\nu\sigma}F^{\beta\nu}V^{\sigma}=-\hbar Q\tilde{F}_{\mu\sigma}V^{\sigma}.
\label{eq:onshell-a}
\end{eqnarray}
where we have neglected the second order term 
$\hbar\epsilon_{\mu\beta\nu\sigma}\nabla^{\nu}(p^{\beta}V^{\sigma})\sim\hbar^{2}$
following the last line of Eqs. (\ref{eq:im}). Here we have used 
$\tilde{F}^{\rho\lambda}=\frac{1}{2}\epsilon^{\rho\lambda\mu\nu}F_{\mu\nu}$. 

From the 2nd, 3rd and 5th lines of Eqs. (\ref{eq:real}), the pseudoscalar,
vector and tensor components are 
\begin{eqnarray}
P & = & -\frac{1}{2m}\hbar\nabla^{\mu}A_{\mu},\nonumber \\
V_{\mu} & = & \frac{1}{m}p_{\mu}F-\frac{1}{2m}\hbar\nabla^{\nu}S_{\nu\mu},\nonumber \\
S^{\nu\sigma} & = & \frac{1}{2m}\hbar(\nabla^{\nu}V^{\sigma}-\nabla^{\sigma}V^{\nu})-\frac{1}{m}\epsilon^{\nu\sigma\alpha\beta}p_{\alpha}A_{\beta}.\label{eq:p-v-s}
\end{eqnarray}
Substituting the above into Eqs. (\ref{eq:onshell-f},\ref{eq:onshell-a}),
we obtain a closed system of on-shell equations for $F$ and $A^{\mu}$
up to $O(\hbar)$. We now collect all equations for $F$ and $A^{\mu}$,
\begin{eqnarray}
p^{\mu}A_{\mu} & = & 0,\nonumber \\
p\cdot\nabla A^{\rho} & = & QF^{\rho\xi}A_{\xi},\nonumber \\
p\cdot\nabla F & = & 0,\nonumber \\
(p^{2}-m^{2})F & = & -\frac{1}{2m}\hbar QF_{\mu\nu}\epsilon^{\mu\nu\alpha\beta}p_{\alpha}A_{\beta},\nonumber \\
(p^{2}-m^{2})A_{\mu} & = & -\frac{1}{m}\hbar Q\tilde{F}_{\mu\sigma}p^{\sigma}F,\label{eq:kinetic}
\end{eqnarray}
which make a closed system of equations for $F$ and $A^{\mu}$ and
can be solved perturbatively in powers of $\hbar$. The last two equations
relate the solutions of the lower order to the higher order. Having
$F$ and $A^{\mu}$, we can determine $P$, $V^{\mu}$ and $S^{\mu\nu}$
through Eq. (\ref{eq:p-v-s}).

\section{Wigner function components at leading order}

\label{sec:lead-order}At leading order of electromagnetic interaction, the gauge link in the Wigner
function in Eq.~(\ref{eq:wigner-def-1}) can be set to 1, then we have
following simple form 
\begin{equation}
W_{\alpha\beta}(x,p)=\int\frac{d^{4}y}{(2\pi)^{4}}e^{-ip\cdot y}\left\langle \bar{\psi}_{\beta}(x+\frac{y}{2})\psi_{\alpha}(x-\frac{y}{2})\right\rangle .\label{eq:wigner-def}
\end{equation}
We can expand fermionic fields in momentum space using creation and
destruction operators as 

\begin{eqnarray}
\psi(x) & = & \frac{1}{\sqrt{\Omega}}\sum_{\mathbf{k},s}\frac{1}{\sqrt{2E_{k}}}[a(\mathbf{k},s)u(\mathbf{k},s)e^{-ik\cdot x}+b^{\dagger}(\mathbf{k},s)v(\mathbf{k},s)e^{ik\cdot x}],\nonumber \\
\bar{\psi}(x) & = & \frac{1}{\sqrt{\Omega}}\sum_{\mathbf{k},s}\frac{1}{\sqrt{2E_{k}}}[a^{\dagger}(\mathbf{k},s)\bar{u}(\mathbf{k},s)e^{ik\cdot x}+b(\mathbf{k},s)\bar{v}(\mathbf{k},s)e^{-ik\cdot x}],\label{eq:field-expansion}
\end{eqnarray}
where $\Omega$ is the volume and $s=\pm$ denote the spin state parallel
or anti-parallel to the spin quantization direction $\mathbf{n}$
in the rest frame of the particle. Insert the above into Eq. (\ref{eq:wigner-def}),
we obtain 
\begin{eqnarray}
W_{\alpha\beta}(x,p) & = & \frac{1}{(2\pi)^{3}}\delta(p^{2}-m^{2})\bigg\{\theta(p^{0})\sum_{s}f_{\mathrm{FD}}(E_{p}-\mu_{s})u_{\alpha}(\mathbf{p},s)\bar{u}_{\beta}(\mathbf{p},s)\nonumber \\
 &  & -\theta(-p^{0})\sum_{s}f_{\mathrm{FD}}(E_{p}+\mu_{s})v_{\alpha}(-\mathbf{p},s)\bar{v}_{\beta}(-\mathbf{p},s)\bigg\},\label{eq:wigner-spinor}
\end{eqnarray}
where we have used $\left\langle a^{\dagger}(\mathbf{p},s)a(\mathbf{p},s)\right\rangle =f_{\mathrm{FD}}(E_{p}-\mu_{s})$
and $\left\langle b^{\dagger}(-\mathbf{p},s)b(-\mathbf{p},s)\right\rangle =f_{\mathrm{FD}}(E_{p}+\mu_{s})$
with the Fermi-Dirac distribution defined by $f_{\mathrm{FD}}=1/(e^{\beta x}+1)$ 
($\beta \equiv 1/T$, $T$ is temperature) 
and $\mu_s$ is the chemical potential for the fermions with spin state $s$.

From Eq. (\ref{eq:wigner-spinor}) we can extract the scalar, vector
and axial vector components by applying Eq. (\ref{eq:fpvas}). We
extract the scalar component as 
\begin{eqnarray}
F_{(0)} & = & \mathrm{Tr}[W]=m\delta(p^{2}-m^{2})V
\end{eqnarray}
where we have used $\bar{u}(\mathbf{p},s)u(\mathbf{p},s)=2m$ and
$\bar{v}(-\mathbf{p},s)v(-\mathbf{p},s)=-2m$, and 
\begin{equation}
V\equiv\frac{2}{(2\pi)^{3}}\sum_{s}\left[\theta(p^{0})f_{\mathrm{FD}}(p_{0}-\mu_{s})+\theta(-p^{0})f_{\mathrm{FD}}(-p_{0}+\mu_{s})\right].\label{eq:vector-const}
\end{equation}
For the vector component, we have 
\begin{eqnarray}
V_{(0)}^{\mu} & = & \mathrm{Tr}[\gamma^{\mu}W]=p^{\mu}\delta(p^{2}-m^{2})V,\label{eq:v0}
\end{eqnarray}
where we have used $\bar{u}(\mathbf{p},s)\gamma^{\mu}u(\mathbf{p},s)=2(E_{p},\mathbf{p})$
and $\bar{v}(-\mathbf{p},s)\gamma^{\mu}v(-\mathbf{p},s)=2(E_{p},-\mathbf{p})$.
For the axial vector component, we obtain 
\begin{eqnarray}
A_{(0)}^{\mu} & = & \mathrm{Tr}[\gamma^{\mu}\gamma^{5}W]\nonumber \\
 & = & m\left[\theta(p_{0})n^{\mu}(\mathbf{p},\mathbf{n})-\theta(-p_{0})n^{\mu}(-\mathbf{p},-\mathbf{n})\right]\delta(p^{2}-m^{2})A,\label{eq:a0-1}
\end{eqnarray}
where we have defined 
\begin{equation}
A\equiv\frac{2}{(2\pi)^{3}}\sum_{s}s\left[\theta(p^{0})f_{\mathrm{FD}}(p_{0}-\mu_{s})+\theta(-p^{0})f_{\mathrm{FD}}(-p_{0}+\mu_{s})\right],\label{eq:axial-const}
\end{equation}
and used $\bar{u}(\mathbf{p},s)\gamma^{\mu}\gamma^{5}u(\mathbf{p},s)=2msn^{\mu}(\mathbf{p},\mathbf{n})$
and $\bar{v}(-\mathbf{p},s)\gamma^{\mu}\gamma^{5}v(-\mathbf{p},s)=2msn^{\mu}(-\mathbf{p},-\mathbf{n})$
with $n^{\mu}(\mathbf{p},\mathbf{n})$ given by 
\begin{eqnarray}
n^{\mu}(\mathbf{p},\mathbf{n}) & = & \Lambda_{\;\nu}^{\mu}(\mathbf{v})n^{\nu}(\mathbf{0},\mathbf{n})=\left(\frac{\mathbf{n}\cdot\mathbf{p}}{m},\mathbf{n}+\frac{(\mathbf{n}\cdot\mathbf{p})\mathbf{p}}{m(m+E_{p})}\right).
\end{eqnarray}
Here $\Lambda_{\;\nu}^{\mu}(\mathbf{v})$ is the Lorentz transformation
for $\mathbf{v}=\mathbf{p}/E_{p}$ and $n^{\nu}(\mathbf{0},\mathbf{n})=(0,\mathbf{n})$
is the 4-vector of the spin quantization direction in the rest frame of the fermion. One can check
that $n^{\mu}(\mathbf{p},\mathbf{n})$ satisfies $n^{2}=-1$ and $n\cdot p=0$,
so it behaves like a spin 4-vector up to a factor of 1/2. For Pauli
spinors $\chi_{s}$ and $\chi_{s^{\prime}}$ in $u(\mathbf{p},s)$
and $v(-\mathbf{p},s^{\prime})$ respectively, we have $\chi_{s}^{\dagger}\boldsymbol{\sigma}\chi_{s}=s\mathbf{n}$
and $\chi_{s^{\prime}}^{\dagger}\boldsymbol{\sigma}\chi_{s^{\prime}}=-s^{\prime}\mathbf{n}$.
We can take the massless limit by setting $\mathbf{n}=\hat{\mathbf{p}}$,
then we have $mn^{\mu}(\mathbf{p},\mathbf{n})\rightarrow(|\mathbf{p}|,\mathbf{p})$
and $mn^{\mu}(-\mathbf{p},-\mathbf{n})\rightarrow(|\mathbf{p}|,-\mathbf{p})$.
This way we can recover the previous result of the axial vector
component for massless fermions \cite{Gao:2012ix,Chen:2012ca}, 
\begin{eqnarray}
A_{(0)}^{\mu} & \rightarrow & \delta(p^{2})\frac{2}{(2\pi)^{3}}p^{\mu}\sum_{s}s\bigg\{\theta(p^{0})f_{\mathrm{FD}}(p_{0}-\mu_{s})+\theta(-p^{0})f_{\mathrm{FD}}(-p_{0}+\mu_{s})\bigg\},
\end{eqnarray}
where $s=\pm$ now denote the right-handed and left-handed fermions.

\section{Axial vector component at next-to-leading order}

\label{sec:first-order}
We start with the solution to the Wigner function
for chiral or massless fermions \cite{Gao:2012ix,Chen:2012ca,Gao:2015zka}.
It is well known that in this case the vector and axial vector components
decouple from the rest of  other components. Their solutions can be recombined
into the chiral components of right-hand and left-hand, 
\begin{eqnarray}
\mathscr{J}_{(0)s}^{\rho}(x,p) & = & p^{\rho}f_{s}\delta(p^{2}),\nonumber \\
\mathscr{J}_{(1)s}^{\rho}(x,p) & = & -\frac{s}{2}\hbar\tilde{\Omega}^{\rho\sigma}p_{\sigma}\frac{df_{s}}{d(\beta p_{0})}\delta(p^{2})-sQ\hbar\tilde{F}^{\rho\lambda}p_{\lambda}f_{s}\frac{\delta(p^{2})}{p^{2}},\label{eq:1st-solution}
\end{eqnarray}
where $s=\pm$ denote right-hand/left-hand helicity, $p_{0}\equiv u\cdot p$,
$\tilde{\Omega}^{\rho\sigma}=\frac{1}{2}\epsilon^{\rho\sigma\mu\nu}\partial_{\mu}(\beta u_{\nu})$,
$\tilde{F}^{\rho\lambda}=\frac{1}{2}\epsilon^{\rho\lambda\mu\nu}F_{\mu\nu}$,
and $f_{s}$ are distribution functions of chiral fermions defined by 
\begin{eqnarray}
f_{s}(x,p) & = & \frac{2}{(2\pi)^{3}}\left[\theta(p_{0})f_{\rm FD}(p_{0}-\mu_{s})+\theta(-p_{0})f_{\rm FD}(-p_{0}+\mu_{s})\right].\label{eq:dist}
\end{eqnarray}
and 
\begin{eqnarray}
\frac{df_{s}}{d(\beta p_{0})} & = & \frac{2}{(2\pi)^{3}}\left[\theta(p_{0})\frac{d}{d(\beta p_{0})}f_{\rm FD}(p_{0}-\mu_{s})-\theta(-p_{0})\frac{d}{d(-\beta p_{0})}f_{\rm FD}(-p_{0}+\mu_{s})\right].
\end{eqnarray}
Note that in the definition of the dual vorticity tensor $\tilde{\Omega}^{\rho\beta}$
in Eq. (\ref{eq:1st-solution}) we have included the factor $\beta=1/T$
inside $\partial_{\mu}$, which is different from the convention (without
such a factor) in Refs. \cite{Gao:2012ix,Chen:2012ca,Gao:2015zka}.
The chiral components in Eq. (\ref{eq:1st-solution}) are related
to the vector and axial vector components by 
\begin{eqnarray}
V^{\rho}(x,p) & = & \mathscr{J}_{+}^{\rho}(x,p)+\mathscr{J}_{-}^{\rho}(x,p),\nonumber \\
A^{\rho}(x,p) & = & \mathscr{J}_{+}^{\rho}(x,p)-\mathscr{J}_{-}^{\rho}(x,p).
\end{eqnarray}

Now we try to extend Eq. (\ref{eq:1st-solution}) to massive fermions.
We recall that the vector and axial vector components at the leading
or zeroth order are given by Eqs. (\ref{eq:v0}) and (\ref{eq:a0-1}), 
\begin{eqnarray}
V_{(0)}^{\mu} & = & p^{\mu}\delta(p^{2}-m^{2})V,\nonumber \\
A_{(0)}^{\mu} & = & m\left[\theta(p_{0})n_{\sigma}(\bar{p},n_{0})-\theta(-p_{0})n_{\sigma}(-\bar{p},-n_{0})\right]\delta(p^{2}-m^{2})A,\label{eq:a0-cov}
\end{eqnarray}
where $V=f_{+}+f_{-}$ and $A=f_{+}-f_{-}$ are given by Eqs. (\ref{eq:vector-const}) and (\ref{eq:axial-const}).
Note that we have written relevant quantities in covariant forms with
fluid velocity: $p_{0}\rightarrow u\cdot p$, $(0,\mathbf{p})\rightarrow\bar{p}^{\alpha}=p^{\alpha}-(u\cdot p)u^{\alpha}$,
$E_{p}=\sqrt{m^{2}-\bar{p}^{2}}=|u\cdot p|$. In particular, we have re-written
$n^{\mu}(\mathbf{p},\mathbf{n})$ and $n^{\mu}(-\mathbf{p},-\mathbf{n})$
from Eq. (\ref{eq:a0-1}) as 
\begin{eqnarray}
n^{\mu}(\mathbf{p},\mathbf{n})\rightarrow n^{\mu}(\bar{p},n_{0}) & = & -\frac{n_{0}\cdot\bar{p}}{m}u^{\mu}+n_{0}^{\mu}-\frac{n_{0\xi}\bar{p}^{\xi}\bar{p}^{\mu}}{m(m+E_{p})},\nonumber \\
n^{\mu}(-\mathbf{p},-\mathbf{n})\rightarrow n^{\mu}(-\bar{p},-n_{0}) & = & -\frac{n_{0}\cdot\bar{p}}{m}u^{\mu}-n_{0}^{\mu}+\frac{n_{0\xi}\bar{p}^{\xi}\bar{p}^{\mu}}{m(m+E_{p})},
\end{eqnarray}
where $n_{0}^{\alpha}=(0,\mathbf{n})$ is the four-vector in the co-moving
frame of the fluid cell and satisfies $n_{0}\cdot u=0$. We now propose the
following form for the axial component at the first order for massive
fermions based on the solution in Eq. (\ref{eq:1st-solution}), 
\begin{eqnarray}
A_{(1)}^{\alpha}(x,p) & = & -\frac{1}{2}\hbar\tilde{\Omega}^{\alpha\sigma}p_{\sigma}\frac{dV}{d(\beta p_{0})}\delta(p^{2}-m^{2})-Q\hbar\tilde{F}^{\alpha\lambda}p_{\lambda}V\frac{\delta(p^{2}-m^{2})}{p^{2}-m^{2}},
\label{eq:a1}
\end{eqnarray}
where the first term is induced by the vorticity. We can check that
the above $A_{(1)}^{\alpha}(x,p)$ satisfies the first and last equation
of (\ref{eq:kinetic}). The kinetic equation, the second equation
of Eq. (\ref{eq:kinetic}), can be imposed for $A_{(1)}^{\alpha}(x,p)$.
We will show in the next section that the axial vector can give the
spin 4-vector, so we can calculate the polarization density from the
vorticity term of $A_{(1)}^{\alpha}(x,p)$ in Eq. (\ref{eq:a1}).

\section{Energy-momentum and spin tensor/vector density from the Wigner function}

\label{sec:spin-tensor}
The symmetrized Lagrange density for a free
Dirac particle is 
\begin{equation}
L=\bar{\psi}(\frac{1}{2}i\gamma^{\mu}\overleftrightarrow{\partial}_{\mu}-m)\psi,
\end{equation}
where $\overleftrightarrow{\partial}=\overrightarrow{\partial}-\overleftarrow{\partial}$.
The energy-momentum tensor can be obtained,
\begin{eqnarray}
T^{\mu\nu} & = & \frac{\partial L}{\partial(\partial_{\mu}\psi)}\partial^{\nu}\psi+\partial^{\nu}\psi^{\dagger}\frac{\partial L}{\partial(\partial_{\mu}\psi^{\dagger})}-g^{\mu\nu}L\nonumber \\
 & = & \frac{1}{2}i\bar{\psi}\gamma^{\mu}\overleftrightarrow{\partial}^{\nu}\psi-g^{\mu\nu}\bar{\psi}(\frac{1}{2}i\gamma^{\mu}\overleftrightarrow{\partial}_{\mu}-m)\psi.
\end{eqnarray}
When taking ensemble average of $T^{\mu\nu}$, we will use the Dirac
equation and assume all fields are on-shell. So we have 
\begin{eqnarray}
\left\langle T^{\mu\nu}(x)\right\rangle  & = & \frac{1}{2}i\left\langle \bar{\psi}(x)\gamma^{\mu}\overleftrightarrow{\partial}_{x}^{\nu}\psi(x)\right\rangle -g^{\mu\nu}\left\langle \bar{\psi}(\frac{1}{2}i\gamma^{\alpha}\overleftrightarrow{\partial}_{\alpha}-m)\psi\right\rangle \nonumber \\
 & = & \int d^{4}pp^{\nu}\mathrm{Tr}(\gamma^{\mu}W)-g^{\mu\nu}\int d^{4}p\left[p_{\mu}\mathrm{Tr}(\gamma^{\mu}W)-m\mathrm{Tr}(W)\right]\nonumber \\
 & = & \int d^{4}pp^{\nu}V^{\mu},
\end{eqnarray}
where we have used $p^{\mu}V_{\mu}=mF$, the first line of Eqs. (\ref{eq:real}) and 
\begin{eqnarray}
W_{\alpha\beta}(x,p) & = & \int\frac{d^{4}y}{(2\pi)^{4}}e^{-ip\cdot y}\left\langle \bar{\psi}_{\beta}(x+\frac{y}{2})\psi_{\alpha}(x-\frac{y}{2})\right\rangle \nonumber \\
\lim_{y\rightarrow0}\partial_{y}^{\mu}\left\langle \bar{\psi}_{\beta}(x+\frac{y}{2})\psi_{\alpha}(x-\frac{y}{2})\right\rangle  & = & \frac{1}{2}\left\langle [\partial_{x}^{\mu}\bar{\psi}_{\beta}(x)]\psi_{\alpha}(x)-\bar{\psi}_{\beta}(x)\partial_{x}^{\mu}\psi_{\alpha}(x)\right\rangle \nonumber \\
 & = & i\int d^{4}pp^{\mu}W_{\alpha\beta}(x,p).
\end{eqnarray}

The spin tensor density is defined by
\begin{eqnarray}
M^{\alpha\beta}(x) & = & \psi^{\dagger}(x)\frac{1}{2}\sigma^{\alpha\beta}\psi(x)
=\frac{1}{2}\mathrm{Tr}\left[\gamma_{0}\sigma^{\alpha\beta}\psi(x)\bar{\psi}(x)\right].
\end{eqnarray}
Taking the ensemble average of the spin tensor, we can also express it
in terms of the Wigner function,
\begin{eqnarray}
\left\langle M^{\alpha\beta}(x)\right\rangle  & = & \frac{1}{2}\lim_{y\rightarrow0}\mathrm{Tr}\left[\gamma_{0}\sigma^{\alpha\beta}\psi(x-\frac{y}{2})\bar{\psi}(x+\frac{y}{2})\right]\nonumber\\
 & = & \frac{1}{2}\int d^{4}p\mathrm{Tr}\left[\gamma_{0}\sigma^{\alpha\beta}W(x,p)\right].
\end{eqnarray}
Then we can define the spin tensor component in the Wigner function
as
\begin{eqnarray}
M^{\alpha\beta}(x,p) & \equiv & \frac{1}{2}\mathrm{Tr}\left[\gamma_{0}\sigma^{\alpha\beta}W(x,p)\right]\nonumber \\
 & = & \frac{1}{2}\left[-\epsilon^{0\alpha\beta\rho}A_{\rho}+ig^{\alpha0}\mathrm{Tr}(\gamma^{\beta}W)-ig^{\beta0}\mathrm{Tr}(\gamma^{\alpha}W)\right],
\end{eqnarray}
where we have used $\gamma^{\mu}\sigma^{\nu\alpha}=i(g^{\mu\nu}\gamma^{\alpha}-g^{\mu\alpha}\gamma^{\nu})+\epsilon^{\mu\nu\alpha\lambda}\gamma^{5}\gamma_{\lambda}$.
If we take $\alpha\beta=ij$ (spatial indices), we have a simple relation
\begin{eqnarray}
M^{ij}(x,p) & = & -\frac{1}{2}\epsilon^{ijk}A_{k}(x,p)=\frac{1}{2}\epsilon^{ijk}A^{k}(x,p),
\end{eqnarray}
where $\epsilon_{ijk}$ is 3-dimensional anti-symmetric tensor. The above property can also be seen by the spatial components of $A^{\mu}(x)$  
\begin{equation}
A^{i}(x)=\bar{\psi}(x)\gamma^{i}\gamma^{5}\psi(x)
=\psi^{\dagger}(x)\gamma^{0}\gamma^{i}\gamma^{5}\psi(x)
=\psi^{\dagger}(x)\Sigma_{i}\psi(x),
\end{equation}
where $\Sigma_{i}=\mathrm{diag}(\sigma_i,\sigma_i)$ with $\sigma_i$ being the Pauli matrices. 
Thus we recognize that $A^{i}(x,p)/2$ corresponds to the spin vector component
of the Wigner function from which we can calculate the polarization density.

\section{Polarization from axial vector component}

\label{sec:polar}

We can now calculate the polarization of
massive fermions from the axial vector component obtained in Section
\ref{sec:spin-tensor}. At the leading order, we can obtain the polarization
density by integrating $A_{(0)}^{\alpha}$ in Eq. (\ref{eq:a0-1})
or Eq. (\ref{eq:a0-cov}) over the 4-momentum, 
\begin{eqnarray}
\Pi^{\alpha}_{(0)}(x) & = & \frac{1}{2}\int d^{4}p\: A_{(0)}^{\alpha}(x,p)\nonumber \\
 & = & \frac{1}{2}m\int\frac{d^{3}p}{(2\pi)^{3}}\frac{1}{E_{p}}\sum_{s}s\left[n^{\alpha}(\bar{p},n_{0})\frac{1}{e^{\beta(E_{p}-\mu_{s})}+1}-n^{\alpha}(-\bar{p},-n_{0})\frac{1}{e^{\beta(E_{p}+\mu_{s})}+1}\right]\nonumber \\
 & = & -\frac{1}{2}u^{\alpha}\int\frac{d^{3}p}{(2\pi)^{3}}\frac{n_{0}\cdot\bar{p}}{E_{p}}\sum_{s}s\left[\frac{1}{e^{\beta(E_{p}-\mu_{s})}+1}-\frac{1}{e^{\beta(E_{p}+\mu_{s})}+1}\right]\nonumber \\
 &  & +\int\frac{d^{3}p}{(2\pi)^{3}}\frac{m}{2E_{p}}\left[n_{0}^{\alpha}-\frac{(n_{0}\cdot\bar{p})\bar{p}^{\alpha}}{m(m+E_{p})}\right]\sum_{s}s\left[\frac{1}{e^{\beta(E_{p}-\mu_{s})}+1}+\frac{1}{e^{\beta(E_{p}+\mu_{s})}+1}\right].\label{eq:polar-1}
\end{eqnarray}
If $\mu_{s}=\mu$ does not depend on $s$, we see immediately that
$\Pi^{\alpha}=0$. In this case the non-vanishing polarization can only
come from the first-order contribution from the vorticity term of $A_{(1)}^{\alpha}(x,p)$
in Eq.~(\ref{eq:a1}), 
\begin{eqnarray}
\Pi^{\alpha}(x) & = &\Pi^{\alpha}_{(1)}(x) = -\frac{1}{4}\int d^{4}p\hbar\tilde{\Omega}^{\alpha\sigma}p_{\sigma}\frac{dV}{d(\beta p_{0})}\delta(p^{2}-m^{2})\nonumber \\
 & = & \frac{1}{2}\int\frac{d^{3}p}{(2\pi)^{3}}\hbar\tilde{\Omega}^{\alpha\sigma}\frac{1}{E_{p}}\left\{ \left.p_{\sigma}\right|_{p_{0}=E_{p}}\frac{e^{\beta(E_{p}-\mu)}}{[e^{\beta(E_{p}-\mu)}+1]^{2}}-\left.p_{\sigma}\right|_{p_{0}=-E_{p}}\frac{e^{\beta(E_{p}+\mu)}}{[e^{\beta(E_{p}+\mu)}+1]^{2}}\right\} \nonumber \\
 & = & \frac{1}{2}\hbar\omega^{\alpha}\int\frac{d^{3}p}{(2\pi)^{3}}\left\{ \frac{e^{\beta(E_{p}-\mu)}}{[e^{\beta(E_{p}-\mu)}+1]^{2}}+\frac{e^{\beta(E_{p}+\mu)}}{[e^{\beta(E_{p}+\mu)}+1]^{2}}\right\} ,
\label{eq:spin-mass}
\end{eqnarray}
where we have removed the spin dependence in the chemical potential,
$\mu_{s}=\mu$, and we have used the fact that the spatial part 
of $p_{\sigma}$ gives vanishing momentum integral. We see that the
polarization density is proportional to the vorticity vector $\omega^{\alpha}=\tilde{\Omega}^{\alpha\sigma}u_\sigma$
and is the sum over contributions from fermions and anti-fermions.

We can also obtain the polarization density from the second (electromagnetic field) 
term of $A_{(1)}^{\alpha}(x,p)$ in Eq. (\ref{eq:a1}), 
\begin{eqnarray}
\Pi_{B}^{\alpha}(x) & = & \frac{1}{2}\hbar Q\int d^{4}p \tilde{F}^{\alpha\lambda}p_{\lambda}V\frac{d}{dp_{0}^{2}}\delta(p^{2}-m^{2})\nonumber \\
 & = & -\frac{1}{4}\hbar Q\int d^{4}p\tilde{F}^{\alpha\lambda}u_{\lambda}\frac{dV}{dp_{0}}\delta(p^{2}-m^{2})\nonumber \\
 & = & \frac{1}{2}\hbar Q\beta B^{\alpha}\int\frac{d^{3}p}{(2\pi)^{3}}\frac{1}{E_{p}}\left\{ \frac{e^{\beta(E_{p}-\mu)}}{[e^{\beta(E_{p}-\mu)}+1]^{2}}-\frac{e^{\beta(E_{p}+\mu)}}{[e^{\beta(E_{p}+\mu)}+1]^{2}}\right\} ,
\label{eq:mag}
\end{eqnarray}
where we have used $\delta^{\prime}(x)=-\delta(x)/x$ and that the
spatial part of $p_{\sigma}$ gives vanishing momentum intergal. Also
we have dropped the complete derivative term which is vanishing at
the boundary in momentum space. 

We see from Eqs. (\ref{eq:spin-mass},\ref{eq:mag}) that 
there is a correspondence between $\Pi^{\alpha}(x)$
from the vorticity and $\Pi_{B}^{\alpha}(x)$ 
from the magnetic field: $E_p\omega^\alpha \leftrightarrow Q\beta B^\alpha$. 
Note that there is a factor $\beta$ in the definition of $\omega^\alpha$, 
$\omega^\alpha\equiv (1/2)\epsilon ^{\alpha\rho\mu\nu}u_\rho\partial_{\mu}(\beta u_\nu)$. 
At zero temperature, the anti-fermion parts in Eqs. (\ref{eq:spin-mass},\ref{eq:mag}) 
are vanishing, the momentum integrals can be carried out analytically 
from the Fermi sphere distribution. The correspondence at zero temperature now becomes 
$\mu\omega^\alpha \leftrightarrow Q\beta B^\alpha$, 
where the $\beta$ factor cancels the one in the definition of $\omega^\alpha$ 
so the correspondence does not have temperature dependence. 
From such a correspondence, we see that $\Pi_B^{\alpha}(x)$ always comes with the charge $Q$ 
while $\Pi ^{\alpha}(x)$ does not, therefore the contributions from fermions and anti-fermions 
in $\Pi^{\alpha}(x)$ have the same sign while they have 
opposite signs in $\Pi_B^{\alpha}(x)$ since fermions and anti-fermions 
carry opposite charges.

In this paper we consider only the polarization induced by the vorticity 
since it lasts longer and is stronger than the magnetic effect 
in later stage of hydrodynamical evolution for massive hadrons.

To estimate the magnitude of $\Pi^{\mu}(x)$ for fermions from Eq.
(\ref{eq:spin-mass}), we can carry out the momentum integral in the co-moving frame. 
After completing the integral over the momentum direction, 
we obtain the spin polarization density 
\begin{eqnarray}
\boldsymbol{\Pi}(x) & = & \hbar\boldsymbol{\omega}\frac{1}{4\pi^{2}}\int_{0}^{\infty}d|\mathbf{p}|\,|\mathbf{p}|^{2}\frac{e^{\beta(E_{p}\mp\mu)}}{[e^{\beta(E_{p}\mp\mu)}+1]^{2}}, 
\label{eq:polar-num}
\end{eqnarray}
for fermions $(-)$ and anti-fermions $(+)$. The particle number density for fermions and anti-fermions is given by 
\begin{eqnarray}
\rho (x) & = & 2\int\frac{d^{3}p}{(2\pi)^{3}}\frac{1}{e^{\beta(E_{p}\mp\mu)}+1}
=\frac{1}{\pi^{2}}\int_{0}^{\infty}d|\mathbf{p}|\frac{|\mathbf{p}|^{2}}{e^{\beta(E_{p}\mp\mu)}+1}.
\label{eq:number-density}
\end{eqnarray}
The integrated polarization per particle $\boldsymbol{\Pi}(x)/\rho (x)$ for fermions or anti-fermions  
can be obtained by completing the momentum integrals in Eqs.~(\ref{eq:polar-num}) 
and (\ref{eq:number-density}). We can also define the unintegrated ones 
with momentum dependence, which is given by the following formula in the comoving frame,   
\begin{equation}
\label{polar-momentum}
\frac{\boldsymbol{\Pi}(x,\mathbf{p})}{\rho(x,\mathbf{p})} = \hbar\frac{\boldsymbol{\omega}}{4}\frac{e^{\beta(E_{p}\mp\mu)}}{e^{\beta(E_{p}\mp\mu)}+1},
\end{equation}
where we have defined $\boldsymbol{\Pi}(x,\mathbf{p})\equiv d\boldsymbol{\Pi}(x)/d|\mathbf{p}|$ 
and $\rho(x,\mathbf{p})\equiv d\rho (x)/d|\mathbf{p}|$. 

At zero temperature, the spin polarization density in (\ref{eq:polar-num}) 
and the particle number density in (\ref{eq:number-density}) 
for anti-fermions are vanishing, and the fermion parts 
can be worked out following the Fermi sphere distribution,    
\begin{eqnarray}
\boldsymbol{\Pi}_{T=0}(x) &=& \frac{1}{4\pi^2}\hbar \beta ^{-1} \boldsymbol{\omega}
\mu\sqrt{\mu^2-m^2}\theta (\mu-m),\nonumber\\
\rho _{T=0}(x) &=& \frac{1}{3\pi^2}(\mu^2-m^2)^{3/2}\theta (\mu-m).
\label{zero-t-pol}
\end{eqnarray}
We can also obtain from Eq. (\ref{eq:mag}) the polarization density 
from electromagnetic fields at zero temperature
\begin{equation}
\boldsymbol{\Pi}_{B,T=0}(x)=\frac{1}{4\pi^2}\hbar Q \mathbf{B}\sqrt{\mu^2-m^2}\theta (\mu-m) .
\end{equation} 
We can see the correspondence between $\boldsymbol{\Pi}_{T=0}(x)$ and  $\boldsymbol{\Pi}_{B,T=0}(x)$ 
is $\mu\boldsymbol{\omega} \leftrightarrow Q\beta \mathbf{B}$. 
The integrated polarization per particle $\boldsymbol{\Pi}(x)/\rho (x)$ for fermions 
at zero temperature has a simple form,
\begin{equation}
\frac{\boldsymbol{\Pi}_{T=0}(x)}{\rho _{T=0}(x)} = 
\frac{3}{4} \hbar\beta^{-1} \boldsymbol{\omega} \frac{\mu}{\mu^2-m^2} \theta (\mu-m), 
\label{zero-t-rho}
\end{equation} 
which is a decreasing functuion of $\mu$. Note that 
the factor $\beta ^{-1}$ in Eqs. (\ref{zero-t-pol}, \ref{zero-t-rho}) 
is to cancel the factor $\beta$ in the definition of $\boldsymbol{\omega}$ 
so that there is no temperature dependence in the results.

\begin{figure}
\includegraphics[scale=0.7]{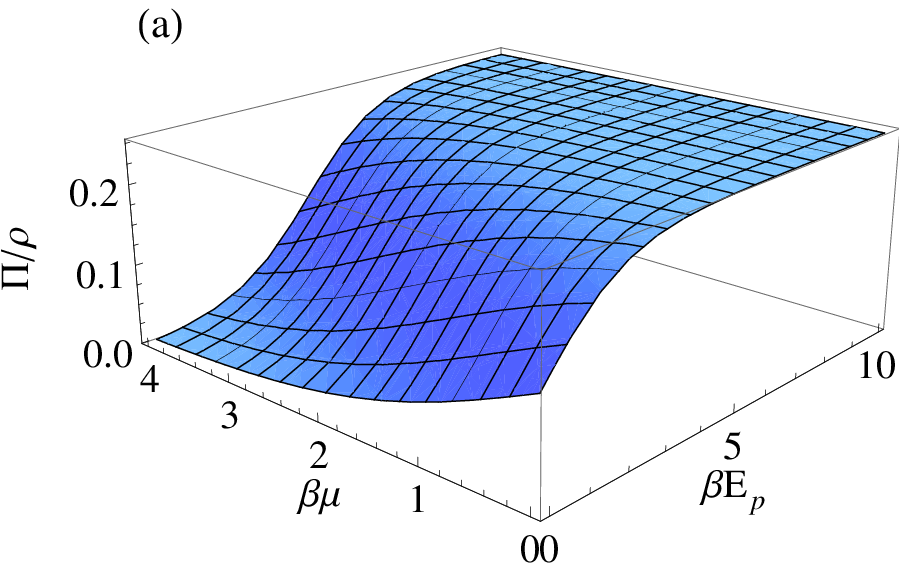}\;\;\;
\includegraphics[scale=0.7]{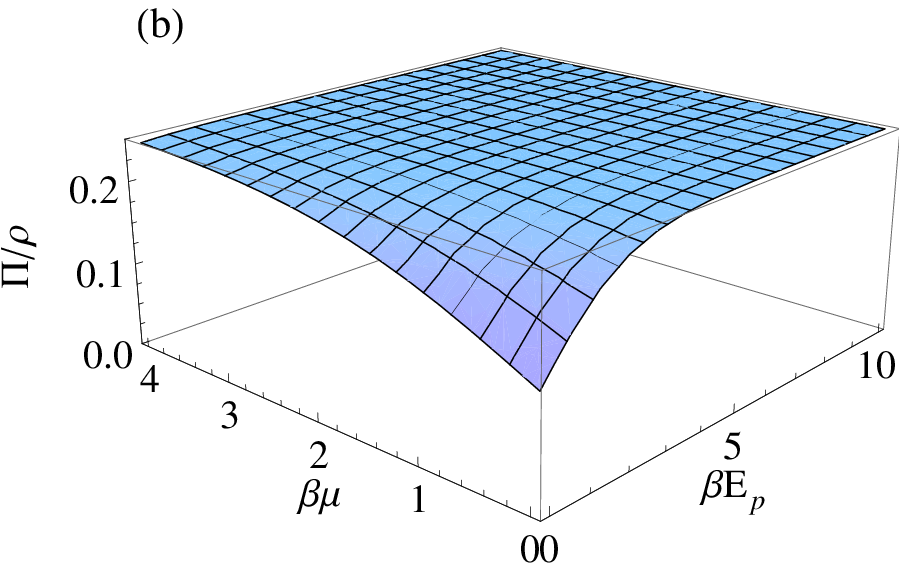}
\caption{\label{fig:pol_em}
The unintegrated polarization per particle defined in Eq.\ (\ref{polar-momentum})  
for fermions (a) and anti-fermions (b) at momentum $\mathbf{p}$  
in the unit of the local vorticity $\hbar\boldsymbol{\omega}$ 
as functions of $\beta E_p$ and $\beta\mu$. }
\end{figure}

\begin{figure}
\includegraphics[scale=0.7]{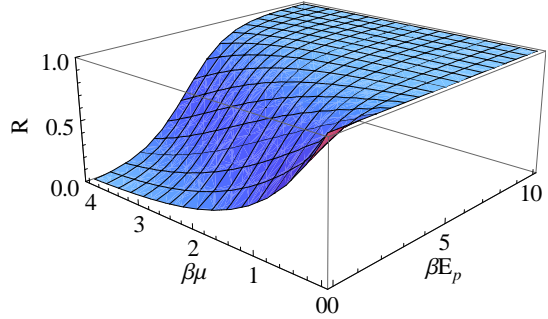}
\caption{\label{fig:pol_em_r}
The ratio $R$ of polarization per particle in Eq.\ (\ref{r-xp})
for fermions to anti-fermions as a function of $\beta E_p$ and $\beta\mu$.}
\end{figure}

The numerical results for the unintegrated polarization per particle 
in Eq.\ (\ref{polar-momentum}) in the unit of the local vorticity 
$\hbar\boldsymbol\omega$ are shown 
in Fig. \ref{fig:pol_em} in the range $\beta E_p=[0,10]$ and $\beta\mu=[0,4]$. 
At fixed values of energy $\beta E_p$, we see that 
$\boldsymbol{\Pi}(x,\mathbf{p})/\rho (x,\mathbf{p})$ 
is a decreasing (increasing) function of $\beta\mu$ 
for fermions (anti-fermions), but it always increases with
$\beta E_p$ at fixed $\beta\mu$ for both fermions and anti-fermions.
The numerical results for the ratio of $\boldsymbol{\Pi}(x,\mathbf{p})/\rho (x,\mathbf{p})$  
for fermions to anti-fermions, 
\begin{equation}
\label{r-xp}
R=\frac{[\boldsymbol{\Pi}(x,\mathbf{p})/\rho(x,\mathbf{p})]_{\mathrm{fermion}}}
{[\boldsymbol{\Pi}(x,\mathbf{p})/\rho(x,\mathbf{p})]_{\mathrm{anti-fermion}}},
\end{equation}
are shown in Fig. \ref{fig:pol_em_r}. We see that $\boldsymbol{\Pi}(x,\mathbf{p})/\rho (x,\mathbf{p})$ 
for fermions is always less than that for anti-fermions, i.e. $R<1$, 
and $R$ decreases with $\beta\mu$ and increases with $\beta E_p$. 
When $\beta E_p$ is very large, the Fermi-Dirac distributions become Boltzmann ones 
and $\boldsymbol{\Pi}(x,\mathbf{p})/\rho (x,\mathbf{p})$ 
reaches its asymptotic value 1/4 (in the unit of $\hbar\boldsymbol\omega$) 
for both fermions and anti-fermions, 
which leads to $R\rightarrow 1$. 

\begin{figure}
\includegraphics[scale=0.7]{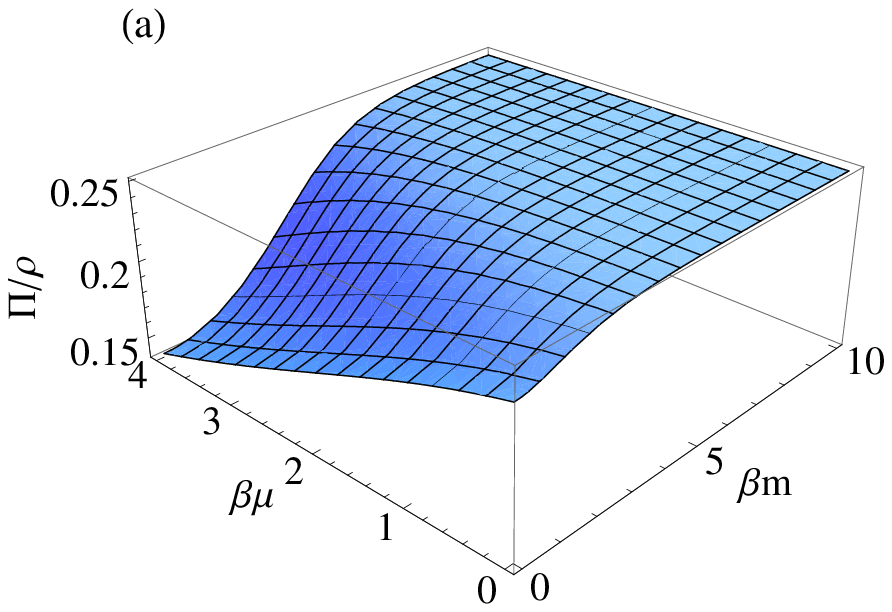}
\includegraphics[scale=0.7]{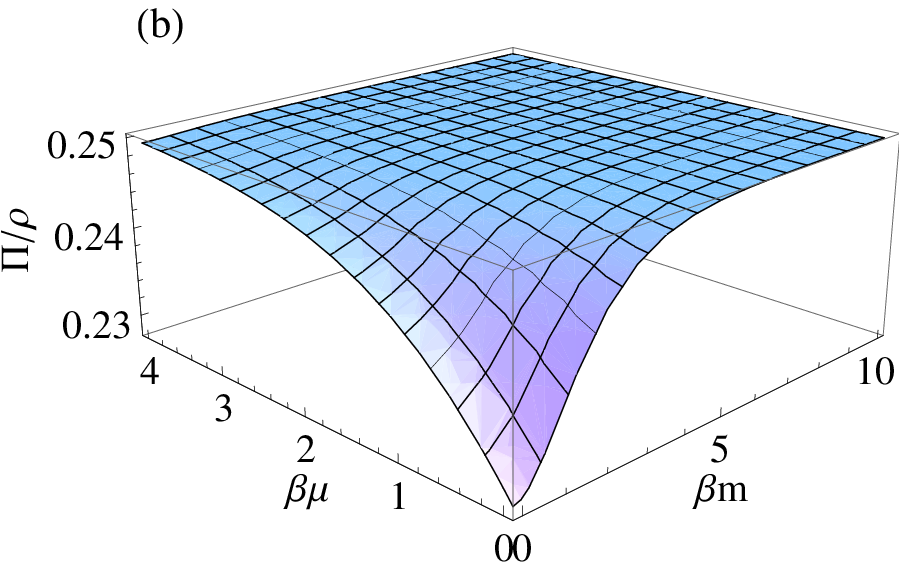}
\caption{\label{fig:pol}
The integrated polarization per particle $\boldsymbol{\Pi}(x)/\rho (x)$
for fermions (a) and anti-fermions (b) in the unit of the local vorticity 
$\hbar\boldsymbol{\omega}$ as functions of $\beta m$ and $\beta\mu$. }
\end{figure}

\begin{figure}
\includegraphics[scale=0.7]{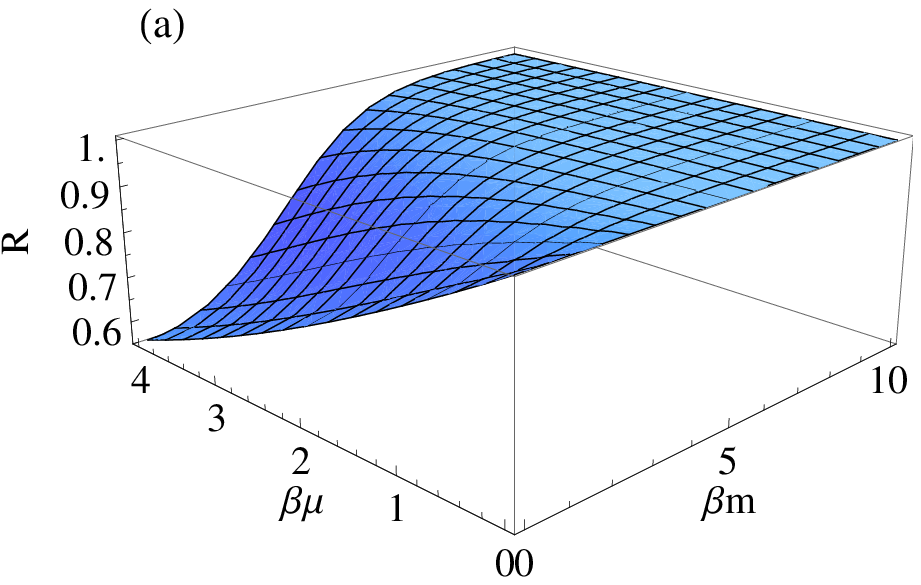}\;\;\;
\includegraphics[scale=0.7]{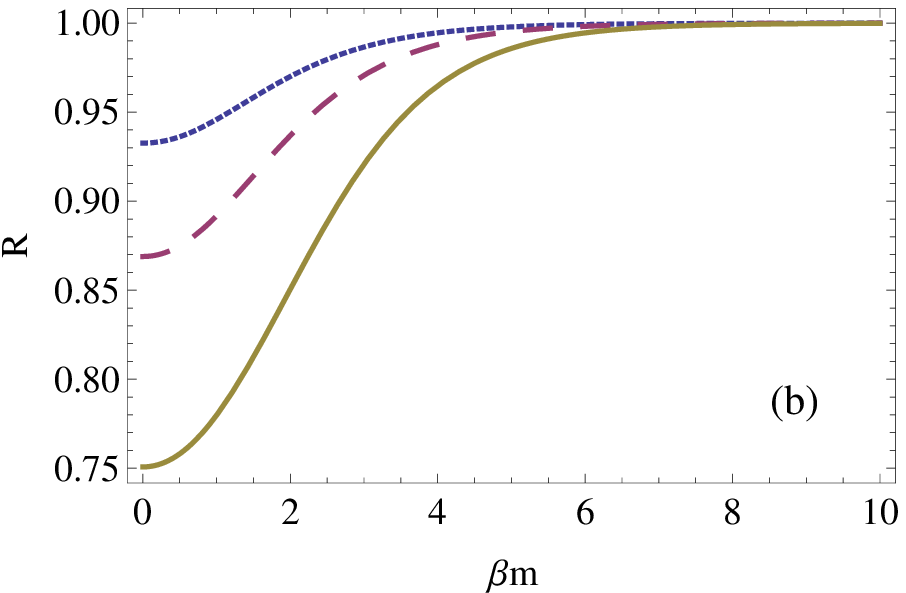}
\caption{\label{fig:ratio-pol}
The ratio $R$ of the integrated polarization per particle in Eq.\ (\ref{r-x})
for fermions to anti-fermions. (a) $R$ 
as a function of $\beta m$ and $\beta\mu$. (b) $R$ as functions
of $\beta m$ at three values $\beta\mu=0.5,\,1,\,2$ corresponding
to short-dashed, long-dashed and solid lines respectively. }
\end{figure}

The numerical results for the integrated polarization per particle $\boldsymbol{\Pi}(x)/\rho (x)$ 
for fermions (left panel) and anti-fermions (right panel) are shown in Fig.\ \ref{fig:pol} as functions
of $\beta m$ and $\beta \mu$. The numerical results for the ratio of 
$\boldsymbol{\Pi}(x)/\rho (x)$, 
\begin{equation}
\label{r-x}
R=\frac{[\boldsymbol{\Pi}(x)/\rho(x)]_{\mathrm{fermion}}}
{[\boldsymbol{\Pi}(x)/\rho(x)]_{\mathrm{anti-fermion}}},  
\end{equation}
are shown in Fig.\ \ref{fig:ratio-pol}. In the left panel we show $R$ 
as a function of $\beta m$ and $\beta \mu$, while in the right panel we show $R$ 
at three values of $\beta \mu$ as functions of $\beta m$. 
The dependences of $\boldsymbol{\Pi}(x)/\rho (x)$ on $\beta m$ 
and $\beta \mu$ are similar to $\boldsymbol{\Pi}(x,\mathbf{p})/\rho (x,\mathbf{p})$ 
on $\beta E_p$ and $\beta \mu$,  
but the variation in the values of $\boldsymbol{\Pi}(x)/\rho (x)$ on $\beta m$ is much smaller 
than $\boldsymbol{\Pi}(x,\mathbf{p})/\rho (x,\mathbf{p})$ 
as shown in Figs.~\ref{fig:pol_em} and \ref{fig:pol_em_r}.  

We see that $R<1$, i.e. the polarization per particle for fermions is always less than that for anti-fermions.
This behavior is consistent to the observation in the STAR experiment \cite{Lisa:2016}.  
Also $R$ decreases with $\mu$ at fixed $m$. Such behaviors are based on the following facts: 
(a) $\boldsymbol{\Pi}(x)$ is actually proportional to the susceptibility $\partial \rho/\partial \mu$ 
and increases/decreases for fermions/anti-fermions with $\beta\mu$ just as $\rho (x)$;
(b) $\boldsymbol{\Pi}_{\mathrm{fermion}}/\boldsymbol{\Pi}_{\mathrm{anti-fermion}}$ 
and $\rho_{\mathrm{fermion}}/\rho_{\mathrm{anti-fermion}}$ are all increasing functions of $\beta\mu$; 
(c) $\boldsymbol{\Pi}_{\mathrm{fermion}}/\boldsymbol{\Pi}_{\mathrm{anti-fermion}}$ 
is less than $\rho_{\mathrm{fermion}}/\rho_{\mathrm{anti-fermion}}$ and 
increases slower with $\beta\mu$ than $\rho_{\mathrm{fermion}}/\rho_{\mathrm{anti-fermion}}$. 

\begin{figure}
\includegraphics[scale=0.7]{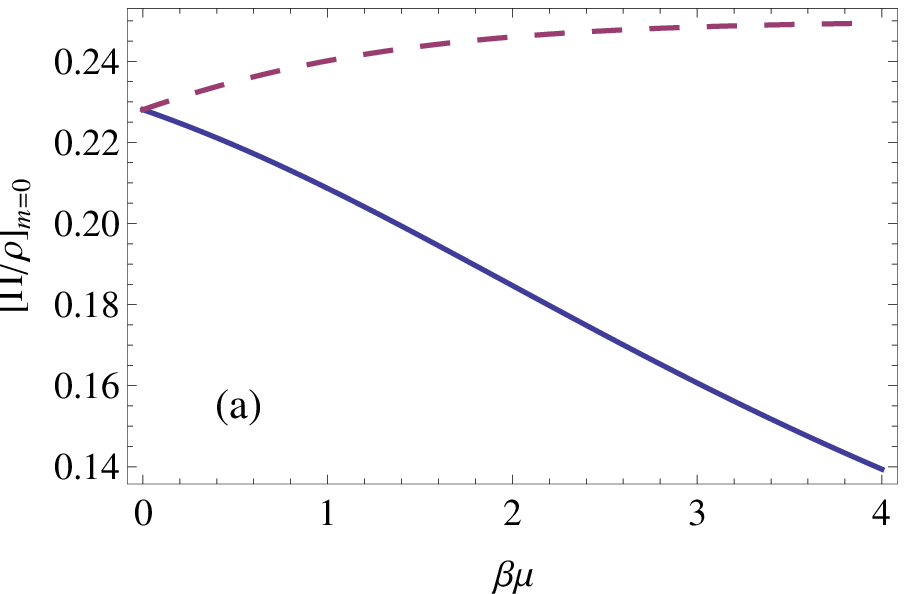}$\;\;\;$\includegraphics[scale=0.7]{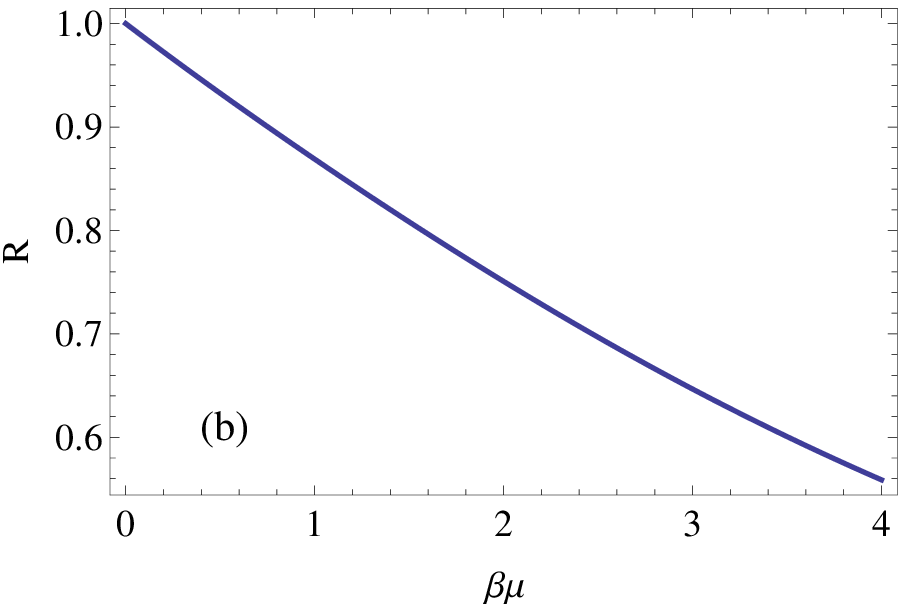}
\caption{\label{fig:massless}
(a) The integrated polarization per particle $\boldsymbol{\Pi}(x)/\rho$
for massless fermions (solid line) and anti-fermions (long-dashed
line) in the unit of  $\hbar\boldsymbol{\omega}$ as functions of $\beta\mu$.
(b) The ratio $R$ of the integrated polarization per particle 
in Eq. (\ref{r-x}) for fermions to anti-fermions as a function of $\beta\mu$. }
\end{figure}

In the massless case, the momentum integrals in Eqs. (\ref{eq:polar-num},\ref{eq:number-density})
can be worked out, so we obtain the quantities for fermions ($+$) and anti-fermions ($-$), 
\begin{eqnarray}
\boldsymbol{\Pi}_{m=0}(x) & = & 
-\hbar\boldsymbol{\omega}\frac{1}{2\pi^{2}}\mathrm{Li}_{2}(-e^{\pm\beta\mu}),\nonumber \\
\rho_{m=0} (x) & = & -\frac{2}{\pi^{2}}\mathrm{Li}_{3}(-e^{\pm\beta\mu}),\nonumber \\
\left[\frac{\boldsymbol{\Pi}(x)}{\rho (x)}\right]_{m=0} & = & \hbar\boldsymbol{\omega}\frac{1}{4}\frac{\mathrm{Li}_{2}(-e^{\pm\beta\mu})}{\mathrm{Li}_{3}(-e^{\pm\beta\mu})},
\end{eqnarray}
where the polylogarithm function is defined by the power series, 
$\mathrm{Li}_{s}(z)=\sum_{k=1}^{\infty}z^{k}/k^{s}$.
Fig. \ref{fig:massless} shows the numerical results for $[\boldsymbol{\Pi}(x)/\rho (x)]_{m=0}$
for fermions and anti-fermions and their ratio $R$ defined by Eq.\ (\ref{r-x})
as functions of $\beta\mu$.  

If we consider the Cooper-Frye description of hadron freezeout in hydrodynamic evolution, 
we can re-write the polarization density in Eq. (\ref{eq:spin-mass})
by replacing the momentum integral with the one on the freezeout hypersurface.
For fermions, we pick up the first term in the second line of Eq. (\ref{eq:spin-mass}) 
and define the polarization spectra in momentum space as,  
\begin{eqnarray}
\label{polar-spec}
\frac{d\Pi^{\alpha}(p)}{d^{3}p} & \approx & \frac{\hbar}{2mE_{p}}\int d\Sigma_{\lambda}p^{\lambda}\tilde{\Omega}^{\alpha\sigma}p_{\sigma}f_{\mathrm{FD}}(x,p)(1-f_{\mathrm{FD}}(x,p)),
\end{eqnarray}
where $p^{\mu}$ denote the on-shell 4-momentum and we have $p^{\mu}=(E_{p},\mathbf{p})$
in the co-moving frame. The particle number distribution for
fermions is given by 
\begin{equation}
f_{\mathrm{FD}}(x,p)=\frac{1}{e^{\beta(x)[u(x)\cdot p-\mu]}+1}.
\end{equation}
In Eq.\ (\ref{polar-spec}), we note that $\Pi^{\alpha}(p)$ is the polarization 
of fermions with the momentum $p$ and has the unit $\hbar$. We can verify that 
the Lorentz transformation rule for both sides of Eq.\ (\ref{polar-spec}) are the same. 
The particle number spectra for fermions in momentum space 
emitting on the freezeout hypersurface can be defined as  
\begin{equation}
\frac{d\rho (p)}{d^{3}p}=\frac{2}{E_{p}}\int d\Sigma_{\lambda}p^{\lambda}f_{\mathrm{FD}}(x,p),
\end{equation}
where the factor 2 is from two spin orientations. Then we obtain the
polarization per particle for fermions with the momentum $p$, 
\begin{eqnarray}
\mathscr{P}^{\alpha}(p) \equiv\frac{d\Pi^{\alpha}(p)/d^{3}p}{d\rho (p)/d^{3}p} & = & \frac{\hbar}{4m}\frac{\int d\Sigma_{\lambda}p^{\lambda}\tilde{\Omega}^{\alpha\sigma}p_{\sigma}\, f_{\mathrm{FD}}(x,p)[1-f_{\mathrm{FD}}(x,p)]}{\int d\Sigma_{\lambda}p^{\lambda}\, f_{\mathrm{FD}}(x,p)}.
\label{eq:freezeout}
\end{eqnarray}
Eq.\ (\ref{eq:freezeout}) is a covariant expression for the polarization vector per particle 
which is the same as the result by Becattini et al \cite{Becattini:2013fla}.
For anti-fermions, we can flip the sign of the chemical potential,
$\mu\rightarrow-\mu$, in the above formula. We see from Eq. (\ref{eq:spin-mass}) 
that the total polarization is the sum of fermion and anti-fermion
contributions.

\section{Summary and conclusion}

We have extended our previous works on the Wigner function 
for chiral or massless fermions to that for massive fermions. 
The Wigner function at the leading order is derived from its definition 
by setting the gauge link to 1 and by expanding the free form of the fermionic quantum
fields in momentum space. Then all components of the Wigner
function can be extracted by projecting the corresponding Dirac matrices and taking
traces. The axial vector component at the next-to-leading order 
for massive fermions can be obtained by extending that for massless fermions 
and satisfies the required equations. 
We have shown that the axial vector component behaves like
a spin 4-vector in phase space up to a factor 1/2. The polarization
density can be computed by integration of the axial vector component
over momentum. Our numerical results show that the polarization per
particle decreases/increases with the (temperature normalized) chemical
potential for fermions/anti-fermions at fixed (temperature normalized)
energy (mass), while it always increases with the (temperature normalized)
energy (mass) at fixed (temperature normalized) chemical potential. We have
found that the polarization per particle for fermions is always less
than that for anti-fermions. At large energy (mass) limit 
the polarization per particle approaches the asymptotic value
$\hbar\boldsymbol\omega/4$ for both fermions and anti-fermions 
following the Boltzmann distribution.
We have also formulated the polarization per particle for fermions with 
the specific momentum on the Cooper-Frye freezeout hypersurface in a hydrodynamic description,  
which is consistent to the previous result of Becattini et al..

\textit{Acknowledgments.} QW is supported in part by the Major State
Basic Research Development Program (MSBRD) in China under the Grant No. 2015CB856902
and 2014CB845406 and by the National Natural Science 
Foundation of China (NSFC) under the Grant No. 11535012. 
XNW is supported in part by the National Natural Science 
Foundation of China (NSFC) under the Grant No. 11221504 and 
by the Chinese Ministry of Science and Technology under Grant No. 2014DFG02050, 
and by the Director, Office of Energy Research, Office of High Energy and Nuclear Physics, 
Division of Nuclear Physics, of the U.S. Department of Energy under Contract No. DE- AC02-05CH11231. 
LGP is supported in part by Helmholtz Young Investigator Group VH-NG-822 
from the Helmholtz Association and GSI.

\bibliographystyle{apsrev}
\addcontentsline{toc}{section}{\refname}\bibliography{ref-1}

\end{document}